\documentclass[]{aastex631}

\usepackage{booktabs}
\usepackage{makecell}
\usepackage{amsmath}

\newcommand{\gaia}{\textit{Gaia}}

\newcommand{\hst}{\textit{HST}}
\newcommand{\ogle}{\textit{OGLE}}

\newcommand{\mic}{$m_{I,\mathrm{syn}}$}
\newcommand{\mih}{$m_{\mathrm{F814W,syn}}$}
\newcommand{\mvj}{$m_{V,\mathrm{syn}}$}

\newcommand{\mio}{$m_{I,\mathrm{OGLE}}$}
\newcommand{\mvo}{$m_{V,\mathrm{OGLE}}$}
\newcommand{\Mic}{$M_{I,\mathrm{syn}}$}
\newcommand{\Mih}{$M_{\mathrm{F814W,syn}}$}

\newcommand{\Mio}{$M_{I,\mathrm{OGLE}}$}

\newcommand{\mi}{$m_I$}
\newcommand{\Mi}{$M_I$}

\newcommand{\Ho}{$H_0$}
\newcommand{\sigs}{$\sigma_s$}
\newcommand{\sigmc}{$\sigma_{\mathrm{MC}}$}
\newcommand{\sigphot}{$\sigma_{\mathrm{phot}}$}
\newcommand{\sigt}{$\sigma_{\mathrm{TRGB}}$}
\newcommand{\Mt}{$M_{\mathrm{TRGB}}$}
\newcommand{\mt}{$m_{\mathrm{TRGB}}$}
\newcommand{\Tcont}{$C_-^+$} %$\mathrm{TRGB}}$}
\newcommand{\Hounits}{$\mathrm{km\,s^{-1}\,Mpc^{-1}}$}

\newcommand{\varstars}{\texttt{SARGs}}
\newcommand{\allstars}{\texttt{Allstars}}
\newcommand{\Aseq}{\texttt{A-sequence}}
\newcommand{\Bseq}{\texttt{B-sequence}}
\newcommand{\ABseq}{\texttt{A\ \& B-sequence}}

\received{December 22, 2023}
\revised{February 6, 2024}
\accepted{February 9, 2024}

\begin{document}

\title{Small amplitude red giants elucidate the nature of the TRGB as a standard candle}

\author[0000-0001-8089-4419]{Richard I. Anderson}
\affiliation{Institute of Physics, \'Ecole Polytechnique F\'ed\'erale de Lausanne (EPFL), Observatoire de Sauverny, Chemin Pegasi 51, 1290 Versoix, Switzerland}

\author[0000-0001-5396-5824]{Nolan W. Koblischke}
\affiliation{Institute of Physics, \'Ecole Polytechnique F\'ed\'erale de Lausanne (EPFL), Observatoire de Sauverny, Chemin Pegasi 51, 1290 Versoix, Switzerland}
\affiliation{Department of Computer Science, Math, Physics, and Statistics, University of British Columbia, Okanagan Campus, 3187 University Way, Kelowna, BC V1V 1V7, Canada}
\affiliation{Department of Astronomy and Astrophysics, University of Toronto, 50 St. George Street, Toronto, ON M5S 3H4, Canada}

\author[0000-0002-0182-8040]{Laurent Eyer}
\affiliation{Department of Astronomy, University of Geneva, Chemin Pegasi 51, 1290 Versoix, Switzerland}

% \collaboration{20}{(AAS Journals Data Editors)}




\begin{abstract}
The tip of the red giant branch (TRGB) is an important standard candle for determining luminosity distances. 
Although several $10^5$ small amplitude red giant stars (SARGs) have been discovered, variability was previously considered irrelevant for the TRGB as a standard candle.
Here, we show that all stars near the TRGB are SARGs that follow several period-luminosity sequences, of which sequence A is younger than sequence B as predicted by stellar evolution. 
We measure apparent TRGB magnitudes, \mt, in the Large Magellanic Cloud (LMC), using Sobel filters applied to photometry from the Optical Gravitational Lensing Experiment and the ESA \gaia\ mission, and we identify several weaknesses in a recent LMC-based TRGB calibration used to measure the Hubble constant. 
We consider four samples: all Red Giants (RGs), SARGs, and sequences A \& B.
The B-sequence is best suited for measuring distances to old RG populations, with $M_{\mathrm{F814W},0} = -4.025 \pm 0.014 \mathrm{(stat.)} \pm 0.033 \mathrm{(syst.)}$\,mag assuming the LMC's geometric distance.
Control of systematics is demonstrated using detailed simulations. 
Population diversity affects \mt\ at a level exceeding the stated precision: the SARG and A-sequence samples yield $0.039$\,mag and $0.085$\,mag fainter (at $5\sigma$ significance) \mt\ values, respectively. Ensuring equivalent RG populations is crucial to measuring accurate TRGB distances.
Additionally, luminosity function smoothing ($\sim 0.02$\,mag) and edge detection response weighting (as much as $-0.06$\,mag) can further bias TRGB measurements, with the latter introducing a tip-contrast relation.
We are optimistic that variable red giants will enable further improvements to the TRGB as a standard candle.
\end{abstract}

\keywords{Standard Candles, Stellar distance, Red giant tip, Pulsating variable stars, Giant stars, Hubble constant, Luminosity function}

% \date{Accepted 9 February 2024}


\section{Introduction} \label{sec:intro}
The extragalactic distance ladder (DL) measures the local Universe expansion rate, or Hubble's constant, $H_0$, using two types of standard candles. In the so-called Hubble flow, type-Ia supernovae (SNeIa) provide precise relative luminosity distances that allow charting cosmic expansion. SNeIa require external luminosity calibration provided by stellar standard candles that in turn are calibrated using geometric distances. The currently most precise stellar standard candles are classical Cepheids \citep{Riess2022H0,CruzReyes2023}, followed by the Tip of the Red Giant Branch (TRGB) \citep{Freedman2021,Scolnic2023}. It has been recently established that $H_0$ is systematically faster than predicted by the best-fit parameters of the $\Lambda$CDM cosmological model adjusted to observations of the early Universe by the ESA mission Planck \citep{Planck,Riess2022H0,DiValentino2021review}. The statistical significance of this \emph{Hubble tension} currently exceeds $5\sigma$ \citep{Riess2022H0,Riess2022Clusters,Murakami2023} when SNeIa are calibrated using Cepheids. A TRGB-based DL calibration by the Carnegie Chicago Hubble Program (CCHP) \citep{Freedman2019,Freedman2021} has yielded a lower $H_0$, compatible with both Cepheids and Planck, inspiring further improvements to the TRGB calibration  \citep{Jang2021N4258,Anand2022,Wu2022,Li2022,Li2023,Scolnic2023}. However, the CCHP's most recent TRGB calibration in the Large Magellanic Cloud (LMC) remains at odds with these improvements \cite[henceforth: H23]{Hoyt2023}. 

The TRGB is an empirical feature of color-magnitude diagrams (CMDs) of old stellar populations and highly useful for determining distances \citep{Lee1993}. Astrophysically, it is ascribed to the luminosity where first ascent low-mass \citep[][$M \lesssim 2.2\,M_\odot$]{Kippenhahn} red giant (RG) stars undergo the Helium flash \citep{Renzini1992}, and stellar models predict a small dependence on age and chemical composition \citep{Salaris2005}. Empirically, the TRGB magnitude is determined using edge detection techniques \citep{Lee1993,Sakai1996} or maximum likelihood analyses \citep{Mendez2002}, preferably in $I-$band where a color dependence is minimal. Two main approaches of correcting age or metallicity effects are used: the CCHP considers the $I-$band luminosity function (LF) to be insensitive to metallicity \citep{Madore2018} and asserts that the blue part of the RG branch select only the oldest stars \citep{Hatt17,Freedman2019}; the Extragalactic Distance Database (EDD) team uses dereddened color to explicitly correct metallicity differences  \citep{Rizzi2007,Anand2021}. Both calibrations agree at $V-I=1.3$ and differ by $\sim 0.10$\,mag at $V-I=1.8$. 

Conceptually, the TRGB distance method compares the inflection points of uni-dimensional LFs composed of generally different (and possibly diverse) RG populations. 
The ability to individually distinguish the desired old and metal-poor RG stars from other stars of similar color and magnitude, such as Asymptotic Giant Branch stars, would improve TRGB distance accuracy. 

RG variability has a long history \citep{Stebbins1930,Fox1982,Edmonds1996}, and wide-spread variability near the TRGB was noted two decades ago \citep{Eyer2002,Ita2002}. Tens of thousands of giants in the Magellanic Clouds follow multiple period-luminosity (PL) sequences labeled A', A, B, C', etc. \citep{Wood1999,Soszynski2004}, and several sequences exhibit clear TRGB features \citep{RedVarsIKiss2003}. Following \citep{Wray2004}, we refer to the small amplitude RG stars with periods of weeks to months  
as \varstars. Although \varstars\ have been used to investigate the RG populations and 3D structure of the Magellanic Clouds \citep{RedVarsIIKiss2004,RedVarsIIILah2005}, the variability of RG stars was not previously considered for measuring distances using the TRGB method. 

Here, we investigate LMC RG star variability aiming to improve the understanding of RG diversity and, in turn, to improve  TRGB calibration and distances. \S\ref{sec:data} describes the observational data sets used, \S\ref{sec:allvar} shows that indeed all stars near the RG tip are measurably variable, and \S\ref{sec:mtrgb} presents our LMC-based measurements of \mt\ and calibration of \Mt. \S\ref{sec:conclusions} summarizes our results and discusses the implications of these result for TRGB distances and \Ho. The appendix provides additional information on sample selection (App.\,\ref{app:cuts}), presents extensive simulations used to investigate TRGB systematics (App.\,\ref{sec:simulations}), and critically compares our results with H23 (App.\,\ref{sec:hoyt}).

\section{Observational data sets used \label{sec:data}}
\subsection{Photometry\label{sec:Photometry}}
We collected \ogle-III $V-$ and $I-$band photometry (\mio, \mvo) from photometric maps \citep{Udalski2008} for \allstars\ and computed mean magnitudes from LPV time-series \citep{Soszynski09} for \varstars\ and \ABseq, cf. Sect.\,\ref{sec:samples} for the sample definitions. The mean uncertainty for \allstars\ was $0.009$\,mag and $0.001$\,mag for the time averages. Color excess uncertainties dominate the total photometric uncertainty.\label{sec:ogle}

Additionally, we collected the following from \gaia\ DR3 \citep{GDR3_summary} table \texttt{gaiadr3.source}: $G-$band magnitudes (\texttt{phot\_g\_mean\_mag}), integrated $G_{RP}$ spectrophotometry (\texttt{phot\_rp\_mean\_mag}), and synthetic photometry from the \gaia\ Synthetic Photometry Catalogue (GSPC) \citep{Montegriffo2022} in Johnson $V-$band (\mvj, \texttt{v\_jkc\_mag}), Cousins $I_c-$band (\mic, \texttt{i\_jkc\_mag}) and \hst\ ACS/WFC F814W (\mih, \texttt{f814w\_acswfc\_mag}).

As shown in the \gaia\ DR3 documentation \citep{Riello2021,Montegriffo2022}, GSPC photometry accurately reproduces photometric standards in the Landolt and Stetson collections (JKC) and \hst\ photometry of MW globular clusters from the HUGS project \citep{HUGS} to within a couple of mmag in the magnitude range of interest ($13 < G < 17$\,mag). We apply zero-point corrections from Tabs.\,2 and G.2 in \cite{Montegriffo2022},  
$\Delta$ \mic $=0.002 \pm 0.014$\,mag
and $\Delta$ \mih $= 0.001 \pm 0.017$\,mag, where $\Delta = m_{\mathrm{std}} - m_{\mathrm{synth}}$ the offset between the reference system and the synthetic photometry. Zero-point uncertainties were added in quadrature to photometric uncertainties. This may slightly overestimate \mih\ uncertainties, since the zero-points for F814W magnitudes were estimated in globular clusters that are more crowded than the LMC. 

Blended stars and foreground objects were removed using a series of astrometric and photometric quality cuts detailed in App.\,\ref{app:cuts}. 
Stars with $\varpi / \sigma_\varpi > 5$, and $\varpi - \sigma_\varpi > 1/50000$\,kpc or proper motions significantly different from the LMC's ($\left( 0.511\cdot( \mu_\alpha^* - 1.88 ) - 0.697\cdot(\mu_\delta - 0.37)\right)^2 + \left( 1.11\cdot(\mu_\alpha^* - 1.88) + 0.4\cdot(\mu_\delta -0.37) \right)^2 < 1$) \citep{GaiaEDR3_LMC} were considered foreground stars if RUWE $< 1.4$. Blending indicators $\beta < 0.99$ \citep[cf. Sec.\,9.3 in][]{Riello2021}, \texttt{ipd\_frac\_multi\_peak} $< 7$, \texttt{ipd\_frac\_odd\_win} $< 7$, and $C^* < 1\sigma$, as well as RUWE $< 1.4$, were used following Sec.\,9.3 in \cite{Riello2021}. 

Direct photometric comparisons reveal excellent agreement between \mio\ and \mic\ (mean and dispersion of $-0.011 \pm 0.023$\,mag for $145937$ RGs, $-0.005 \pm 0.026$\,mag for $30\,315$ \varstars, $-0.002\pm 0.035$\,mag for $15965$ stars close to the TRGB), and our extinction-corrected TRGB measurements in these passbands also agree to within $1-4$\,mmag.
\mio\ and \mih\ also agree extremely well with mean and dispersion $0.002 \pm 0.028$\,mag for \allstars\ and $0.008 \pm 0.026$\,mag for \varstars.

\subsection{Reddening and extinction\label{sec:red}}
We corrected reddening using color excess values, $E(V-I)$, based on LMC Red Clump stars \citep{Skowron2021ApJS}. Individual uncertainties on $E(V-I)$ and the systematic reddening error of $0.014$\,mag are included in the analysis (Sect.\,\ref{meth:errors}). Removing stars with $E(V-I) > 0.2$ limits 
the effect of reddening law uncertainties to $\lesssim 0.02$\,mag. 

We computed reddening coefficients, $R_I$, for each passband using \texttt{pysynphot} \citep{pysynphot} for an $R_V = 3.3$ recalibrated Fitzpatrick reddening law \citep{Fitzpatrick1999,Schlafly2011} and a star near the TRGB following \cite{Anderson2022} and obtained $R_{I_c} = 1.460$, $R_{\mathrm{ACS,F814W}}=1.408$, $R_{G_{RP}}=1.486$, and $R_{\mathrm{Gaia,G}}=1.852$, consistent with $R_I \sim 1.5$ used to construct the reddening maps \citep{Skowron2021ApJS}. Analogously, we determine $R^W_{VI}=1.457$ to compute reddening-free Wesenheit magnitudes $W_{VI}= I - R^W_{VI}\cdot (V-I)$ for fitting and inspecting SARG PL sequences.

\subsection{Sample selections\label{sec:samples}\label{sec:allstars}}
We selected four samples for TRGB measurements and refer to them as \allstars, \varstars, and \ABseq. All samples were cleaned using objective photometric and astrometric data quality indicators, cf. App.\,\ref{app:cuts}.  
The \allstars\ sample consists of all stars from the \ogle-III LMC photometric maps \citep{Udalski2008}. \gaia\ source\_ids were cross-matched using the positions from \ogle, cf. App.\,\ref{app:cuts} for details. We also considered sample selections following H23 in App.\,\ref{sec:hoyt}.

\label{sec:varstars}
The \varstars\ sample is a subset of \allstars\ and consists of $40\,185$ Small Amplitude Red Giants stars from the \ogle-III catalog of variable stars \citep{Soszynski09} cross-matched with \gaia. 
\label{sec:sequences}
The \ABseq\ samples are distinct subsets of \varstars\ selected in two steps. Each was first crudely selected by manually defining a region around each sequence in the reddening-free (Wesenheit) magnitudes, $W_{VI} = I - 1.457\cdot(V-I)$, versus $\log{P}$ diagram. In turn, we fitted quadratic P-W relations to the initial selections and applied a $3\sigma$ outlier rejection to obtain $W_{VI} = 12.434 - 4.582\cdot \log(P_1/20\mathrm{d}) - 2.098\cdot (\log_{10}(P_1/20\mathrm{d}))^2 $ for the A-sequence, and $W_{VI} = 12.290 - 5.096\cdot \log_{10}(P_1/35\mathrm{d}) - 2.414\cdot (\log(P_1/35\mathrm{d}))^2 $ for the B-sequence. All periods discussed refer to the dominant period, $P_1$, despite \varstars\ being multi-modal pulsating stars. The adopted pivot periods are close to the sample medians.

\section{All stars at the RGB tip are variable \label{sec:allvar}}

\begin{figure}[h]
    \centering
    \includegraphics[]{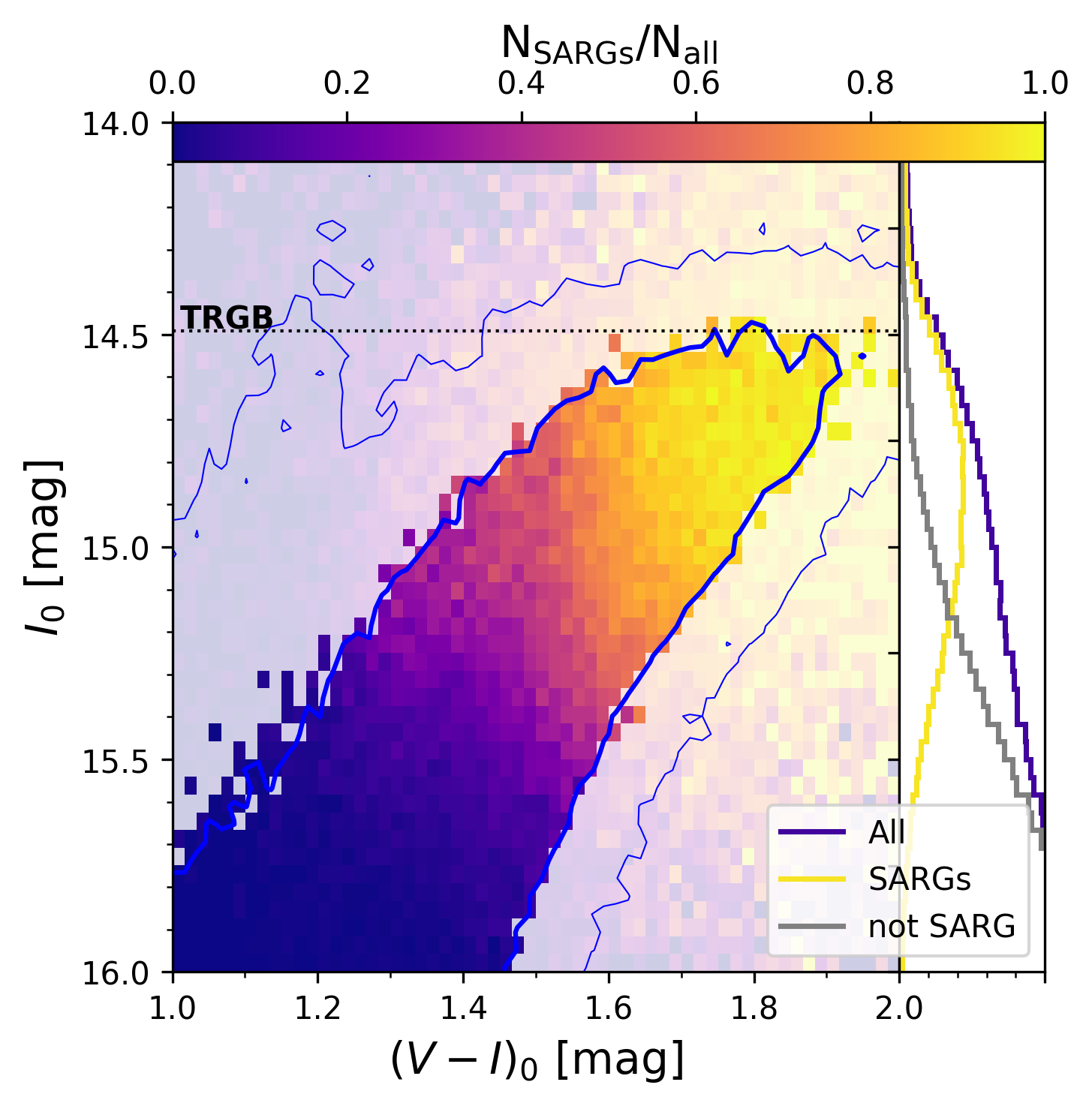}
    \caption{Virtually all stars at the TRGB are variable. \textit{Left:} Coarsely-binned CMD indicating the fraction of \varstars\ to all stars using a color map indicated at the top. Blue contours highlight the RGB's location by delineating the $70\%$ (thicker) and $20\%$ (thinner) contours of all stars in the \ogle-III LMC photometric map \citep{Udalski2008,Soszynski09}. The TRGB's location is shown as a dotted line. The fraction of variable stars increases upward towards the TRGB, reaching nearly $100\%$ at the tip. \textit{Right:} LFs of \allstars\ (black), \varstars\ (orange), and their complement (gray).}
    \label{fig:CMD}
\end{figure}

Figure\,\ref{fig:CMD} shows for the first time that virtually all LMC stars at the TRGB are \varstars, and that the fraction of \varstars\ decreases toward fainter magnitudes. This is potentially significant, as stars near the TRGB are typically regarded as non-variable \citep[e.g.][]{Madore2023}. Given small amplitudes of typically $0.01-0.04$\,mag, and long-period multi-periodicity, this variability is mostly negligible for distance measurements. However, we show below that \varstars\ variability offers useful new insights into the populations of RG stars near the tip.

\subsection{Understanding \varstars\label{sec:varcause}}
SARGs are high-luminosity RGs that maintain self-excited multi-mode non-radial pulsations of low longitudinal degree and low radial order $n$. The pulsational nature has been established based on observed period ratios \citep{Wood2015}, and non-linear non-adiabatic pulsation models \citep{Takayama2013,Trabucchi2017,Xiong2018,McDonald2019}. SARG pulsations are higher-overtone, lower-amplitude analogs to semi-regular and Mira variables \citep{Xiong2018,Trabucchi2021SRV} and can be distinguished from stochastic solar-like oscillations seen in lower-luminosity red giants in the period-amplitude plane \citep{TaburPLRMgiants2010,Banyai2013,Auge2020}. PL sequences defined by the dominant pulsation mode $P_1$ represent an evolutionary timeline, whereby stars first evolve to higher luminosity along PL sequence A until the mode of the next lower radial order becomes dominant. Once the dominant pulsation mode changes, the star continues its ascent to higher luminosity on the next longer-period (lower $n$) sequence B \citep{Wood2015,Trabucchi2019}. 

\subsection{Dissecting population diversity using RG variability\label{sec:diversity}}
The evolutionary scenario suggests that the older and younger RG populations can be directly separated via their PL sequences. This would be very useful because PL sequences employ properties observable in the individual stars of interest. We investigated this possibility using stellar ages determined by spectroscopy from the Apache Point Observatory Galaxy Evolution Experiment (APOGEE) data release 17 \citep{Povick2023}. 

\begin{figure*}
    \centering
    \includegraphics[width=1\textwidth]{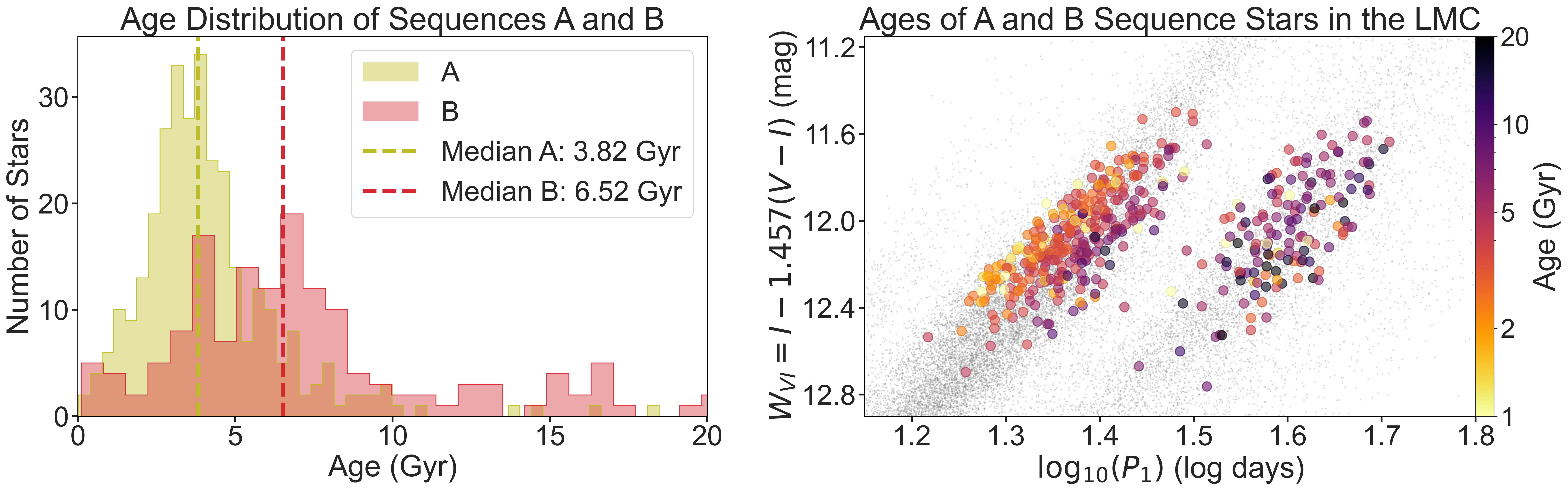}
    \caption{Ages of \varstars\ from \cite{Povick2023}, based on APOGEE near-infrared spectra. \textit{Left:} Histogram of stellar ages on SARG Sequences A and B. Sequence B is significantly older than Sequence A, as well as less metallic (not shown). The ages of the oldest stars are particularly uncertain. \textit{Right:} Period-Wesenheit relations of \ogle-III SARGs near Sequences A and B with ages color-coded. Sequence A exhibits significant age differences from its short period boundary to the long-period boundary, consistent with an evolutionary sequence. Stars on sequence B are significantly older.\label{fig:Povick}}
\end{figure*}

Figure\,\ref{fig:Povick} shows that the \Bseq\ indeed contains the older (median age $6.5$\,Gyr), lower-metallicity RG stars near the tip. Thus, \Bseq\ stars correspond most closely to the astrophysically motivated target RG population near the He flash. \Aseq\ stars are systematically younger ($3.8$\,Gyr) and exhibit a trend of increasing stellar age with period. Additionally, \Aseq\ stars are more metal-rich and of higher mass than \Bseq\ stars. This empirically confirms the theoretical evolutionary scenario of the LPV PL sequences proposed by \cite{Wood2015}. We note that the ratio of stars on the two sequences is relatively even across the \ogle-III LMC footprint, with mean $N_A / N_B = 2.26 \pm 0.03$, cf. App.\,\ref{sec:hoyt}. Hence, the ratio of older to younger stars appears mostly constant within this area, although age and metallicity gradients are known at larger angular separations \citep{Majewski2009,Munoz2023}. Figure\,\ref{fig:CMD2} shows that the A \& B-sequence samples fully overlap in the color-magnitude diagram, so that variability information is required to separate the samples. Furthermore, we note that all \varstars\ PL sequences are shifted to shorter periods in the SMC \citep{Soszynski2011}, which indicates the lower metallicity of RGs in the SMC (Koblischke \& Anderson, in prep.).

\subsection{On the applicability of geometric corrections to LMC RG stars\label{sec:geometry}}

Geometric corrections are commonly applied to stellar populations in the LMC. For example, \citet{Pietrzynski19} provided a geometric model based on detached eclipsing binaries, \citet{Cusano_21} based on RR~Lyrae stars, and \citet{Choi2018} using red clump stars from the Survey of the MAgellanic Stellar History \citep[SMASH]{SMASHDR2}. Previously, the spatial distribution of RG stars in the LMC has been reported to resemble a classical halo \citep{Majewski2009}. In the context of TRGB measurements, H23 discussed the LMC's geometry at length and applied a tailored correction. 

An important validation of the applicability of geometric corrections to classical Cepheids is that they reduce the dispersion of the observed Leavitt laws introduced by the inclination along the line of sight \citep[e.g.][]{Breuval2022,Bhuyan2023}. In analogy to this, we investigated the effect of different geometric corrections on the PL scatter of the \ABseq\ samples. Using the reddening-free $P-W$ sequence fits to the \ABseq\ samples, all geometric corrections reported in the literature increase the rms of the P-W sequence residuals. Specifically, geometric corrections from the literature increase rms from originally $0.148$\,mag / $0.202$\,mag (\Aseq\ / \Bseq) to the following values: $0.153 / 0.207$\,mag \citep{Pietrzynski19}, $0.156/0.209$\,mag \citep{Cusano_21}, and $0.158/0.211$\,mag (H23). Hence, the dispersion of the A \& B-sequence P-W relations do not indicate a need for applying geometric corrections to the RG stars inside the entire \ogle-III footprint, although they also do not strongly rule them out. While we do not apply any geometric corrections to our samples, we note that this has a negligible ($< 0.005$\,mag) effect on our results that are based on the full \ogle-III footprint.

\section{Calibrating the TRGB in the LMC using \varstars\label{sec:mtrgb}}

\subsection{Tip measurement methodology\label{meth:sobel}}
\begin{figure}
    \centering
    \includegraphics[width=1\textwidth]{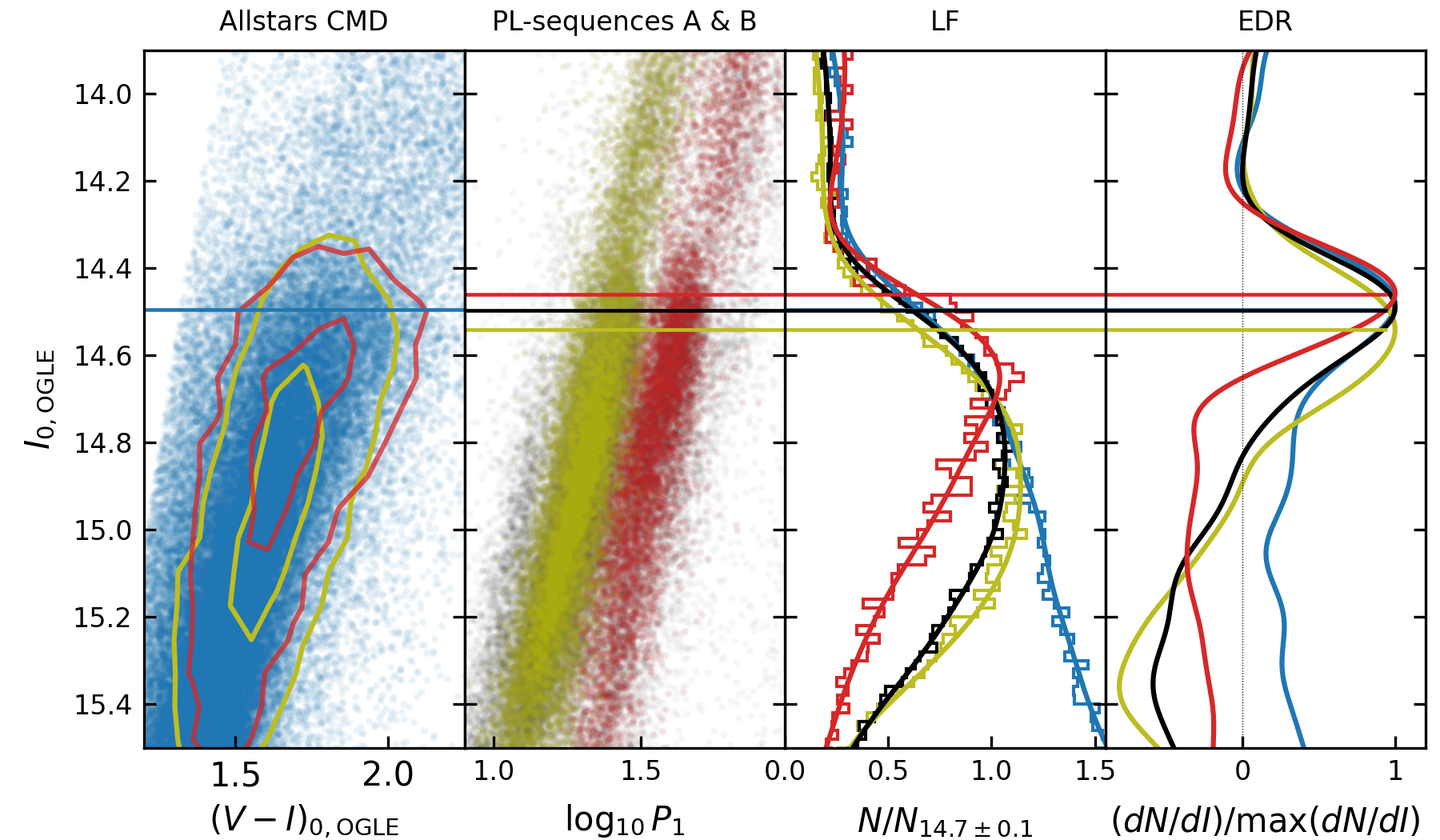}
    \caption{Illustrations of LFs and samples. Horizontal lines show \mio\ for the four samples, \allstars\ (blue), \varstars\ (black), \Aseq\ (yellow), and \Bseq\ (red). From left to right: CMD of the \allstars\ sample (cf. Tab.\,\ref{tab:samples}) with contours (95 and 99 percentiles) of the A and B sequences overlaid; \ogle-III \varstars\ (black points) in the vicinity of the A and B PL-sequences shown as yellow and red points; binned and smoothed LFs (here: \sigs$=0.15$\,mag) normalized to the number of stars within $0.1$\,mag of \mio$=14.7$\,mag; Sobel filter response curves normalized to peak height. Variability information is required to distinguish the A \& B sequences from other RGs and SARGs, since the samples fully overlap in color-magnitude space.}
    \label{fig:CMD2}
\end{figure}

Following the CCHP approach, we measured \mt\ as the inflection point of smoothed and binned (bin size $0.002$\,mag) luminosity functions (LFs) using a [-1,0,1] Sobel filter \citep{Hatt17}. Five sets of photometric measurements were considered, one from \ogle\ and four from \gaia\ DR3 (Sect.\,\ref{sec:Photometry}). Smoothing was applied using a Gaussian-windowed LOcal regrESSion (GLOESS) algorithm \citep{Persson04} as parameterized by \sigs\ to reduce noise \citep{Sakai1996} at the penalty of introducing correlation among magnitude bins. Although smoothing introduces possible bias \citep{Cioni2000,Hatt17}, this is a standard practice. In SNeIa galaxies, the optimal value of \sigs\ is determined using artificial star tests performed on the observed frames \citep{Hatt17}. However, this approach cannot be applied to \ogle\ and \gaia\ photometric catalogs. To remedy this situation, we specifically considered the effects of smoothing as a systematic uncertainty by computing \mt\ for a large range of \sigs\ values and adopting an objective criterion for selecting the acceptable range (see below). 

Table\,\ref{tab:errors} details the determination of statistical and systematic uncertainties. For each value of \sigs, the statistical uncertainty \sigt\ was determined using 1000 Monte Carlo resamples, whereby stellar magnitudes were randomly offset according to their total photometric uncertainties, \sigphot, that combine in quadrature the uncertainties on mean magnitudes, $E(V-I)$ (generally dominant), and the dispersion of the \gaia\ spectrophotometric standardization, cf. Sec.\,\ref{sec:Photometry}. The average \sigphot$=0.088$\,mag is dominated by the color excess uncertainty. Although bootstrapping would yield smaller dispersion (by approx. 2/3), we prefer MC resampling because it allows to consider individual photometric and extinction uncertainties. 
 
We calculated the median \mi\ together with the standard deviation for each computed value of \sigs. The final value of \mt\ for a given stellar sample and photometric data is calculated from the lowest continuous range of \sigs\ values, where $\vert d m_{\mathrm{TRGB}} / d \sigma_s \vert \le 0.1$, that is, where \mt\ is insensitive to the choice of \sigs. Specifically, we adopted the median \mt\ over this \sigs\ range as the final center value result for a given stellar sample and photometric data set. As Fig.\,\ref{fig:mtrgb_smooth} shows, the smallest \sigs\ value is close to \sigphot. Extending the range to higher \sigs\ values is acceptable when the result is not altered by the smoothing, and our automated approach thus allows us to benefit from the enhanced precision afforded by higher \sigs\ \citep{Hatt17} while avoiding bias. For a given sample, the total uncertainty on \mt, \sigt, sums in quadrature the median MC error, the dispersion of \mt, and the dispersion of the difference between mean and median \mt\ in the stable \sigs\ range.

\begin{table}
\centering
\begin{tabular}{@{}llcc@{}}
   Uncertainty  & Includes / based on & \\
   \midrule 
   Statistical \\
   \midrule 
    \sigphot\ & \multicolumn{3}{l}{Average: $0.088$\,mag. Quadratic sum of} \\
    & \multicolumn{3}{l}{individual photometric uncertainties} \\
    & \multicolumn{3}{l}{\citep{Udalski2008,Soszynski09,Montegriffo2022}}, \\
    & \multicolumn{3}{l}{$R_I \times \sigma_{EVI}$ (dominates \sigphot),  $\sigma_{EVI}$ is mean of $\sigma_1$, $\sigma_2$ from \cite{Skowron2021ApJS}} \\
    & \multicolumn{3}{l}{\gaia\ GSPC: standardization uncertainties from \cite{Montegriffo2022}}\\
    \sigmc & \multicolumn{3}{l}{Dispersion of 1000 MC resamples per \sigs\ value} \\
    \sigt & \multicolumn{3}{l}{Reported in Tab.\,\ref{tab:results}}. \\
     & \multicolumn{3}{l}{Sums quadratically across \sigs\ range, where $\vert dm_I / d\sigma_s\vert \le 0.1$:} \\
    & \multicolumn{3}{l}{median \sigmc,} \\
    & \multicolumn{3}{l}{standard deviation of \mt\ values,} \\
    & \multicolumn{3}{l}{difference between mean and median \mt} \\
   \midrule
   Systematic & & \mt\ & \Mt \\
   \midrule
   bin size \& phase  &  variations & 0.001\,mag & 0.001\,mag \\
   $E(V-I)$ & systematic uncertainty from \cite{Skowron2021ApJS} & $0.014$\,mag & $0.014$\,mag \\
   Method bias & \varstars\ sample (App.\,\ref{sec:simulations}) & $0.009$\,mag & $0.009$\,mag \\
    & \allstars\ sample (App.\,\ref{sec:simulations}) & $0.013$\,mag & $0.013$\,mag \\
   \ogle-III zero-point  & from \cite{Udalski2008} & & $0.010$\,mag$^\dagger$ \\
   LMC geometry & impact of corrections on \mt & 0.005\,mag & 0.005\,mag \\
   LMC distance & total uncertainty from \cite{Pietrzynski19} &  & $0.027$\,mag$^\ddagger$ \\
   \midrule
\end{tabular}
\caption{Overview of uncertainties and their determination. \sigphot\ differs for every star and is dominated by the uncertainties of the reddening correction. \sigmc\ is determined per smoothing value (\sigs) via MC resampling. \sigt\ is the final statistical uncertainty reported for \mt. Systematic uncertainties are reported separately if not already included in \sigphot. $^\dagger$: applies only to \Mio. $^\ddagger$: applies only to absolute magnitudes. Effects of population diversity are not listed here, since they depend on many factors, including observer choices. Studies aiming to measure TRGB distances should carefully consider their possible impact, which can be on the order of $0.04$\,mag, the difference between \Bseq\ and \varstars. \label{tab:errors}}
\end{table}

Weighting schemes commonly applied to the edge detection responses (EDRs) produced by the Sobel filter are discussed in detail in App.\,\ref{sec:TCR}, where we show using simulations that EDR weighting biases \mt. Depending on the LF shape and the type of EDR weighting introduced, bias of up to $-0.02$\,mag (SNR weighting) and $-0.07$\,mag (Poisson weighting) can be incurred. Furthermore, we show that this \emph{weighting bias} explains the empirically determined tip-contrast-relations reported based on weighted EDR TRGB magnitudes \citep{Wu2022,Li2023,Scolnic2023}. Since unweighted EDRs yield virtually unbiased results (less than $0.005$\,mag for \Bseq, $0.010$\,mag for \varstars), we prefer the simpler unweighted EDR. However, we caution that comparing our LMC calibration to TRGB magnitudes measured using weighted EDRs must be corrected for these systematic differences.

The systematic uncertainty due to bin size and phase is $0.001$\,mag. Additional systematic uncertainties were assessed using extensive simulations presented in App.\,\ref{sec:simulations}.

\subsection{TRGB apparent magnitudes and uncertainties\label{meth:errors}}
\label{sec:smoothing}

\begin{figure}
    \centering
    \includegraphics[width=1\textwidth]{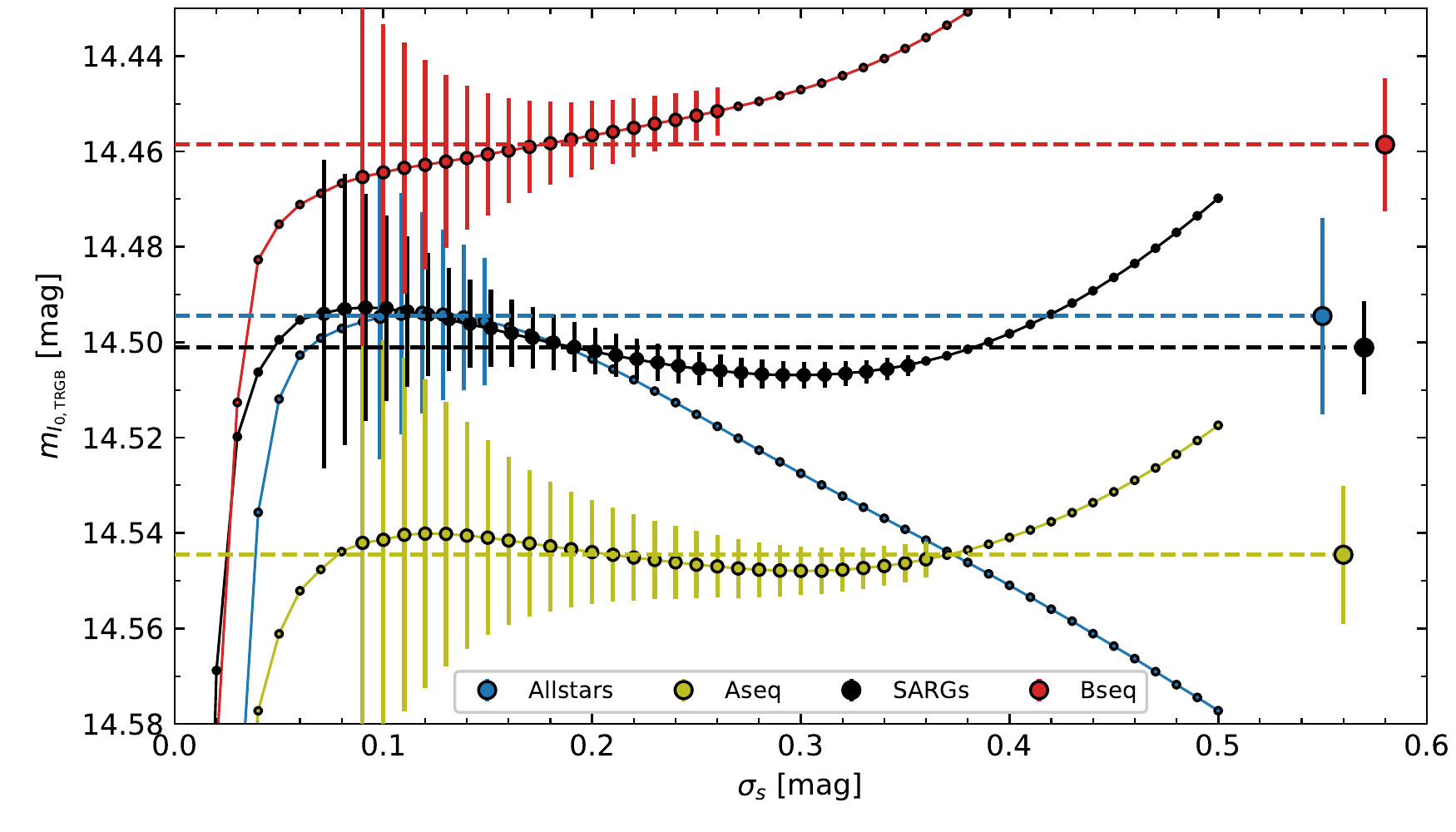}
    \caption{Dependence of \mi\ on \sigs. Errorbars are plotted for the \allstars, \varstars, Aseq, and Bseq samples across the range of \sigs\ values deemed stable, which is narrow for \allstars\ and wide for \varstars. Results from Tab.\,\ref{tab:results} are shown as large errorbars on the right.}
    \label{fig:mtrgb_smooth}
\end{figure}

\begin{table}[]
    \centering
    \begin{tabular}{@{}l@{\hskip 2mm}c@{\hskip 2mm}c@{\hskip 2mm}c@{\hskip 2mm}c@{\hskip 2mm}c@{\hskip 2mm}c}
       & $\langle \mathrm{(V-I)}_0\rangle$ & $I_{\mathrm{OGLE},0}$ & $I_{\mathrm{syn},0}$ & $\mathrm{F814W}_{\mathrm{syn,0}}$ & $G_{{RP},0}$ & $G_0$ \\
sample     & (mag) & (mag) & (mag) & (mag) & (mag) & (mag) \\
     \midrule
\varstars  & $1.82 \pm 0.19$ & $14.501 \pm 0.010$ & $14.497 \pm 0.011$ & $14.491 \pm 0.010$ & $14.640 \pm 0.009$ & $15.621 \pm 0.015$ \\
\allstars    & $1.78 \pm 0.21$ & $14.495 \pm 0.021$ & $14.478 \pm 0.029$ & $14.473 \pm 0.027$ & $14.635 \pm 0.024$ & $15.618 \pm 0.014$ \\
\midrule
\allstars$^\dagger$ & $1.76 \pm 0.11$ & $14.527 \pm 0.027$ & $14.506 \pm 0.035$ & $14.499 \pm 0.032$ & $14.648 \pm 0.039$ & $15.626 \pm 0.036$ \\
Aseq   & $1.79 \pm 0.15$ & $14.545 \pm 0.013$ & $14.543 \pm 0.012$ & $14.537 \pm 0.012$ & $14.690 \pm 0.009$ & $15.655 \pm 0.016$ \\
Bseq   & $1.82 \pm 0.19$ & $14.459 \pm 0.014$ & $14.457 \pm 0.015$ & $14.452 \pm 0.013$ & $14.607 \pm 0.013$ & $15.566 \pm 0.030$ \\
\midrule
    \end{tabular}
    \caption{TRGB apparent magnitudes in the LMC, corrected for extinction. $I_{\mathrm{syn},0}$ and $\mathrm{F814W}_{\mathrm{syn,0}}$ denote synthetic photometry based on \gaia\ DR3 XP spectra (Sect.\,\ref{sec:Photometry}). $G_{{RP},0}$ and $G_0$ are integrated spectrophotometric and white-light magnitudes from \gaia. Results reported are the medians across the range where $\vert dm_{\mathrm{TRGB}} / d\sigma_s\vert \le 0.1$. Uncertainties (\sigt) combine in quadrature the median MC error, \mt\ dispersion, and the dispersion of the difference between \mt\ values determined as means and medians from the MC samples. Additional detail on uncertainties is found in Sec.\,\ref{meth:errors} and Tab.\,\ref{tab:errors}. Median $\mathrm{(V-I)_0}$ colors within $\pm 0.10$\,mag of \mt\ are reported with the $16 - 84$ percentile range. $^\dagger$: color cut following H23.}
    \label{tab:results}
\end{table}

Figure\,\ref{fig:mtrgb_smooth} shows the values of \mt\ obtained for the four stellar samples using \ogle\ $I-$band photometry over a large range of \sigs\ values, and Tab.\,\ref{tab:results} tabulates the results for all samples and photometric data sets. Unsurprisingly, we find significant correlation between \mt\ and \sigs. Surprisingly, however, this correlation depends strongly on LF shape: the \varstars\ and \Aseq\ samples exhibit only weak dependence on \sigs, followed by \Bseq. Smoothing bias is strongest for the \allstars\ sample, for which only a small range of \sigs\ is acceptable according to our criterion. The appropriate ranges for \varstars\ and \allstars\ are $0.07 \le \sigma_s \le 0.36$ and $0.10 \le \sigma_s \le 0.15$, respectively, cf. Fig.\,\ref{fig:mtrgb_smooth}. The total systematic uncertainty on \mt\  is $0.017$\,mag and $0.019$\,mag for the \varstars\ and \allstars\ samples, respectively (cf. Tab.\,\ref{tab:errors}).

The most striking feature of Fig.\,\ref{fig:mtrgb_smooth} and Tab.\,\ref{tab:results} is the significant difference in \mt\ determined from the four stellar samples. The \Bseq\ yields the brightest TRGB measurement in all samples and for all \sigs\ values, $-0.021$\,mag brighter than \varstars, and $-0.085$\,mag brighter than \Aseq, consistent with 
the evolutionary paradigm of the LPV sequences \citep{Wood2015,Trabucchi2019}. While the difference between the \ABseq\ samples is significant at the $5\sigma$ level according to the stated uncertainties, we note that the real significance of this result is much greater due to shared systematics that overestimate the uncertainties in this direct comparison. We also point out that the \Bseq\ is significantly brighter than the \Aseq, independently of \sigs. Despite the \Bseq\ LF's stronger AGB contamination, it is both more concentrated and steeper near \mt\ (cf. Fig.\,\ref{fig:CMD2}) than the other LFs. Last, but not least, we caution against restrictive color cuts for TRGB measurements, since \Bseq\ stars are on average redder than \Aseq\ stars, despite being more metal-poor. A color selection intended to preferentially select metal-poor giants (as applied in the CCHP analysis) would thus unfortunately remove the \Bseq\ stars closest to \mt, resulting in the opposite of the desired effect. However, the degree to which \Bseq\ stars would be removed will be both a function of the metallicity of the RG stars and the expected contamination by \Aseq\ stars, which may be expected to be lower in SNIa host galaxy halos.

The \varstars\ sample yields the most precise TRGB calibration ($\sim 0.5\%$ in distance, not considering uncertainty due to absolute calibration) thanks to the insensitivity to smoothing bias. Our \varstars\ results is also fully consistent with our \allstars\ result, as expected, since \varstars\ contain nearly $100\%$ of the stars in the \allstars\ sample near the RGB tip. The \Aseq\ and \Bseq\ samples yield \mt\ of similar precision, each to about $0.6\%$ in distance. 

Although the \varstars\ result lies close to the middle between \Aseq\ and \Bseq, we stress three important points: a) the \varstars\ sample contains nearly $100\%$ of all RG stars near the Tip (Fig.\,\ref{fig:CMD}); b) the \varstars\ sample contains $\sim 10\,500$ additional stars that are not part of the \ABseq\ samples (notably at longer $P_1$); c) the mean value of \mt\ determined from different samples is generally not identical to the value of \mt\ determined from the combination of the samples (see, e.g., H23, App.\,\ref{sec:hoyt}).

\subsection{Absolute TRGB calibration\label{sec:absolute}}

Our measured apparent TRGB magnitudes can be translated to absolute magnitudes using the LMC's distance derived from detached eclipsing late-type binaries \citep[$\mu = 18.477 \pm 0.006 (\mathrm{stat})\pm 0.026 (\mathrm{syst})$]{Pietrzynski19}. For the sake of comparing to \Ho\ measurements in the literature, we here consider results based on ACS/F814W synthetic photometry from \gaia\ DR3 \citep{Riello2021,Montegriffo2022} as our baseline.

The \varstars\ sample yields \mih$=14.491 \pm 0.010 (\mathrm{stat}) \pm 0.017 (\mathrm{syst})$\,mag, which becomes the most accurate ($1.5\%$) TRGB calibration to date, with \Mih$= -3.986 \pm 0.010 (\mathrm{stat}) \pm 0.033 (\mathrm{syst})$\,mag. This result agrees closely with three LMC-independent TRGB calibrations, including \cite{Li2023} based on the megamaser host galaxy NGC\,4258 ($-3.980 \pm 0.035$\,mag after considering differences in EDR weighting and the appropriate tip contrast $R=3.3$, cf. Sec.\,\ref{sec:TCR}), the EDD TRGB calibration \citep[\Mic$= -4.00 \pm 0.03$\,mag at $(V-I)_0 = 1.82$\,mag]{Rizzi2007}, and calibrations based on \gaia\ parallaxes \citep{Li2023edr3}. 

The \Bseq\ sample yields a slightly ($-0.039$\,mag) brighter calibration of \Mih$=-4.025 \pm 0.014 (\mathrm{stat}) \pm 0.033 (\mathrm{syst})$\,mag ($1.7\%$ in distance), which remains consistent to within $< 1\sigma$ with the aforementioned LMC-independent TRGB calibrations.

\section{Conclusions and Implications for TRGB distances and the Hubble constant\label{sec:conclusions}}

We showed that all stars near the RGB tip in the LMC are variable and found that the \varstars\ LF yields a more accurate TRGB measurement than the \allstars\ sample. We confirmed the evolutionary scenario of small amplitude red giants that the \Bseq\ should contain older stars than the \Aseq\ using spectroscopically determined ages from \citet{Povick2023}. As expected from theory, the oldest stars (\Bseq) yield the brightest tip measurement. Both calibrations support the value of \Mt\ used by \citet{Scolnic2023}, who measured $H_0 = 73 \pm 2$\,\Hounits\ in full agreement \footnote{As discussed by \citet{Scolnic2023}, proper treatment of SNeIa and their peculiar velocities raises \Ho\ more substantially than the calibration of \Mt} with the SH0ES distance ladder calibrated using Cepheids \citep{Riess2022H0}. Both samples yield \Mi\ fainter than used by the CCHP analysis \citep{Freedman2021}, implying higher \Ho. Furthermore, we identified several shortcomings in the LMC-based TRGB calibration by H23 (cf. App.\,\ref{sec:hoyt}), which carries the highest weight in the CCHP \Ho\ analysis. We demonstrate the robustness of our results using extensive simulations.

We elucidated several systematics of the TRGB methodology, notably including \emph{smoothing bias} and (for the first time) \emph{EDR weighting bias}, both of which can impact TRGB distances at the level of a few percent. Specifically, Poisson EDR weighting \citep{Madore2009} can easily bias \mt\ bright by approx. $-0.04$ to $-0.06$\,mag, whereas smoothing can cause bias on the order of $0.02$\,mag depending on LF shape. We therefore strongly recommend that TRGB analyses  always specify how smoothing was treated, and we introduced the smoothing slope as an objective criterion for selecting an appropriate range of \sigs. We furthermore stress that TRGB distances based on weighted EDRs must be standardized using tip-contrast-relations \citep{Wu2022} that arise due to EDR weighting bias (App.\,\ref{sec:TCR}). This applies in particular to CCHP TRGB distances determined using weighted EDRs. We attribute the insensitivity of the \varstars\ and \ABseq\ samples to smoothing bias to the shape of their observed LFs, which fall off towards fainter magnitudes, similar to LFs of RG populations in SNeIa host galaxies, cf. Fig.~3 in \cite{Freedman2019}. The impact of photometric incompleteness on LF shape, which depends on many factors and varies among observed fields, deserves additional attention, notably with regards to smoothing and EDR weighting bias.

Our results highlight population diversity as a major concern for measuring distances using the TRGB method. Without spectroscopic information, it is difficult to know the makeup of the RG population under study, and the statistical nature of using the TRGB CMD feature as a standard candle prevents individual standardization and calibration as applied to individual standard candles, such as classical Cepheids. A related issue is field placement for TRGB studies \citep[e.g.,][]{Beaton2018review,Scolnic2023}, which can lead to significant problems for distance determination \citep[e.g.,][]{Csoernyei2023}. As a result, it is not fully clear whether the \varstars\ or \Bseq\ calibration more directly applies to the RG populations in the SN-host galaxies targeted for measuring \Ho\ \citep{Freedman2021,Anand2021,Scolnic2023}. Distances to old metal-poor halos of distant galaxies that are void of intermediate-age RG populations should be based on the \Bseq\ calibration. However, the \varstars\ calibration may be more suitable if the presence of younger stars cannot be excluded. In case the makeup of the RG population cannot be reliably ascertained, then adopting an additional systematic uncertainty of, say, $\sim 0.04$\,mag ($\sim 2\%$ in distance) for TRGB distances appears to be a prudent choice. Interestingly, we find that color cuts targeting the blue RG stars in the hopes of selecting older, metal-poorer stars have the opposite effect when dealing with mixed-age RG populations, since the older population is slightly redder near the Tip. 

Despite these concerns, there is reason for optimism that the TRGB method can be further improved, notably for use with {\it JWST}. The multi-periodic long-period variability of stars near the RGB tip offers promising new avenues for understanding RG population diversity and its impact on distance measurements. As a first step in this direction, Koblischke \& Anderson (in prep.) are targeting \varstars\ in the Small Magellanic Cloud. While the low-level variability of \varstars\ will not be detectable at large distances, we are optimistic that RG variability will inform TRGB standardization using more readily accessible observables.

\begin{acknowledgments}
The authors thank the \ogle\ and \gaia\ collaborations for producing data sets of exceptional quality without which this work would not have been possible. We thank Marc Pinsonneault for detailed comments that helped improve the manuscript. We also thank Joshua Povick for providing results from \cite{Povick2023} ahead of publication. We thank the anonymous referee for their constructive report.

RIA is funded by the SNSF via the Eccellenza Professorial Fellowship PCEFP2\_194638. NWK acknowledges funding through the EPFL Excellence Research Internship Program, the ThinkSwiss Research Scholarship, and the UBCO Go Global Program. 

This research was supported by the International Space Science Institute (ISSI) in Bern, through ISSI International Team project \#490, SHoT: The Stellar Path to the Ho Tension in the Gaia, TESS, LSST and JWST Era. This research made use of the NASA Astrophysics Data System.
\end{acknowledgments}

\vspace{5mm}
\facilities{Gaia, OGLE}

\software{astropy \citep{2013A&A...558A..33A,2018AJ....156..123A}, pandas \citep{pandas}, matplotlib \citep{matplotlib}          }

\pagebreak
\appendix

\section{Sample selection and quality cuts\label{app:cuts}}

The specific ADQL query for retrieving the \ogle-III LMC stars from the \gaia\ archive was the following:

\begin{quote}
\tt
SELECT * \\
FROM gaiadr3.gaia\_source \\
INNER JOIN gaiadr3.synthetic\_photometry\_gspc as S \\
ON S.source\_id = gaia\_source.source\_id\\
WHERE \\
CONTAINS(\\
	POINT('ICRS',gaiadr3.gaia\_source.ra,gaiadr3.gaia\_source.dec),\\
	CIRCLE('ICRS',$80.9$,$-69.75$,$5.8$)\\
)=1\\
AND S.i\_jkc\_mag $> 12.5$ \\
AND S.i\_jkc\_mag $< 16.5$ \\
AND (S.v\_jkc\_mag - S.i\_jkc\_mag) $> 1.0$ \\
AND (S.v\_jkc\_mag - S.i\_jkc\_mag) $< 3.5$ \ .
\end{quote}

Table\,\ref{tab:samples} details the impact of various quality cuts applied to the initial data set as described in \S\ref{sec:data}. 

\begin{table}[h]
    \centering
      \begin{tabular}{@{}llrrr@{}}
        Sample & step & $N_{\mathrm{removed}}$ & $N_{\mathrm{remain}}$ & $N_{\mathrm{applies}}$\\
        \midrule
         & \gaia\ cone search (rough color \& mag cut) & & 540\,596 \\
         & in \ogle-III footprint, have \ogle\ $V-$band & 192\,057 & 348\,539 & \\
         & Foreground stars & 18\,973 & 329\,566 & 18\,973\\
         & LMC proper motion ellipse & 1\,578 & 327\,988 & 20\,371\\
         & $\beta \leq 0.99$ & 87\,322 & 240\,666 & 90\,386\\
         & \texttt{ipd\_frac\_multi\_peak} $< 7$ & 7\,220 & 233\,446 & 31\,246\\
         & \texttt{ipd\_frac\_odd\_win} $< 7$ & 48 & 233\,398 & 382\\
         & $RUWE < 1.4$ & 4\,430 & 228\,968 & 37\,568\\
         & $C^* < 1\sigma$ & 60\,416 & 168\,552 & 147\,175\\
         & $E(V-I) \leq 0.2$ & 12\,459 & 156\,093 & 23\,593\\
         \allstars & Removing stars much bluer than RGB & 15\,319 & 140\,774 & 43\,285\\
         \allstars$^{\dagger}$ & color cuts according to \cite{Hoyt2023} & 24\,683 & 116\,091 & \\
         \midrule
        & \ogle-III SARG catalog \citep{Soszynski09} &  & 79\,200 \\
         & \gaia\ cross-match & 1559 & 77\,641 \\
        \varstars & quality cuts as applied to \allstars &  37\,456 & 40\,185 &  \\
        \Aseq & LPVs on PL sequence A (manual selection) &  & 20\,470 \\
        \Bseq & LPVs on PL sequence B (manual selection) &  & 9\,164 \\
        \midrule
    \end{tabular}
    \caption{Description of sample selections. Columns $N_{\mathrm{removed}}$, $N_{\mathrm{removed}}$, $N_{\mathrm{applies}}$ list the number of stars removed sequentially, the number of stars remaining in the sample after applying the cut, and the number of initial stars to which this cut applies, respectively.
    Foreground and proper motion cuts, as well as the photometric quality indicators are explained in the text.}
    \label{tab:samples}
\end{table}

\section{Simulating luminosity functions to quantify systematics\label{sec:simulations}}
We investigated systematics for the \varstars, \Bseq, and \allstars\ samples using analytic representations of the observed LFs to simulate how variations in LF shape or the measurement process affect \mt. Each LF is composed of three components carefully matched to reproduce the observed LFs: a) an AGB population ($A(m)$) modeled as a Gaussian of height $a_{\mathrm{AGB}}$, dispersion $\sigma_{\mathrm{AGB}}$, and mean offset $\Delta m_{\mathrm{AGB}}$ from the input TRGB magnitude $m_{\mathrm{TRGB}}$; b) a sigmoid function $S(m)$ centered on \mt\ to represent the TRGB ``break'' with sharpness parameter $w$; c) an RG population ($R(m)$) at magnitudes fainter than \mt, which increases rapidly for the \allstars\ sample and decreases otherwise. For \varstars\ and \ABseq\ LFs, we modeled $R(m)$ using a Gaussian whose peak of amplitude $a_{\mathrm{SARG,mode}}$ coincides with the sigmoid function at the mode of the observed LF $m_{\mathrm{SARG,mode}}$. 

The simulated LF is the sum of its components:
\begin{equation}
    N(m) = A(m) + S(m) + R(m)    
\end{equation}
where
\begin{equation}
    A(m) = a_{\mathrm{AGB}} \cdot e^{ - \left ( \frac{m - (m_{\mathrm{TRGB}} - \Delta m_{\mathrm{AGB}}))}{2 \sigma_{\mathrm{AGB}}} \right)^2} \ ,
\end{equation}
\begin{equation}\label{eq:sigmoid}
    S(m) = \frac{1}{1 + e^{\left(-\dfrac{m-m_{\mathrm{TRGB}}}{w}\right)}} ,
\end{equation}
and $R(m)$ is defined relative to $m_{\mathrm{SARG,mode}} = m_{\mathrm{TRGB}} + \Delta m_{\mathrm{SARG}}$ so that
\begin{equation}
    R(m < m_{\mathrm{SARG,mode}})  = 0  
\end{equation}
for both samples, 
\begin{equation}\label{eq:Rall}
    R(m \ge m_{\mathrm{SARG,mode}}) = s_1 (m - m_{\mathrm{SARG,mode}} )^2 + s_2 ( m - m_{\mathrm{SARG,mode}} ) 
\end{equation}
for \allstars, where $s_1$ and $s_2$ are slopes to reproduce the rising \allstars\ LF, and 
\begin{equation}\label{eq:Rosarg}
    R(m \ge m_{\mathrm{SARG,mode}})  = a_{\mathrm{SARG,mode}} \cdot \left ( e^{-\left( \frac{m-m_{\mathrm{SARG,mode}}}{2 \sigma_{\mathrm{SARG,decl}}}\right)^2 } - 1 \right ) 
\end{equation}
for \varstars, with $\sigma_{\mathrm{SARG,decl}}$ the dispersion that reproduces the declining \varstars\ LF towards fainter magnitudes. The baseline parameters for LF simulations are shown in Tab.\,\ref{tab:LF_coeffs}. 

Anchoring Eqs.\,\ref{eq:Rall} and \ref{eq:Rosarg} to $m_{\mathrm{SARG,mode}}$ allows for a common and straightforward definition of TRGB contrast:
\begin{equation}\label{eq:contrast}
    C^+_- = a_{\mathrm{SARG,mode}} / a_{\mathrm{AGB}} \ .
\end{equation}
\Tcont\ is directly measurable for LPV LFs, independently of \mt. Conversely, the tip contrast ratio $R = N_+ / N_-$ \citep{Wu2022} requires knowledge of \mt\ to estimate the number of stars $0.5$\,mag below and above the tip, $N_+$ and $N_-$ (cf. Sect.\,\ref{sec:TCR}). 

\begin{table}[]
    \centering
    \begin{tabular}{@{}lcccccccc@{}}
        sample & $a_{\mathrm{AGB}}$ & $\sigma_{\mathrm{AGB}}$ & $\Delta m_{\mathrm{AGB}}$ & $w$ & $\Delta m_{\mathrm{SARG}}$ & $\sigma_{\mathrm{SARG,decl}}$ & $s_1$ & $s_2$ \\
        \midrule
        \allstars & 0.24 & 0.21 & 0.450 & 0.095 & 0.305 & $-$ & 0.10 & 0.65 \\
        \varstars & 0.20 & 0.21 & 0.450 & 0.080 & 0.305 & 0.31 & $-$ & $-$ \\
        \Bseq &     0.25 & 0.20 & 0.475 & 0.060 & 0.125 & 0.30 & $-$ & $-$ \\
        \Aseq &     0.14 & 0.20 & 0.490 & 0.100 & 0.400 & 0.25 & $-$ & $-$ \\
        \midrule
    \end{tabular}
    \caption{Baseline simulation parameters determined to visually match the observed LF, normalized to unit height at $m_{\mathrm{SARG,mode}}$. $m_{\mathrm{TRGB},0} = 14.495$ was used in Fig.\,\ref{fig:simLF}. }
    \label{tab:LF_coeffs}
\end{table}

\begin{figure}
    \centering
    \includegraphics[width=1\textwidth]{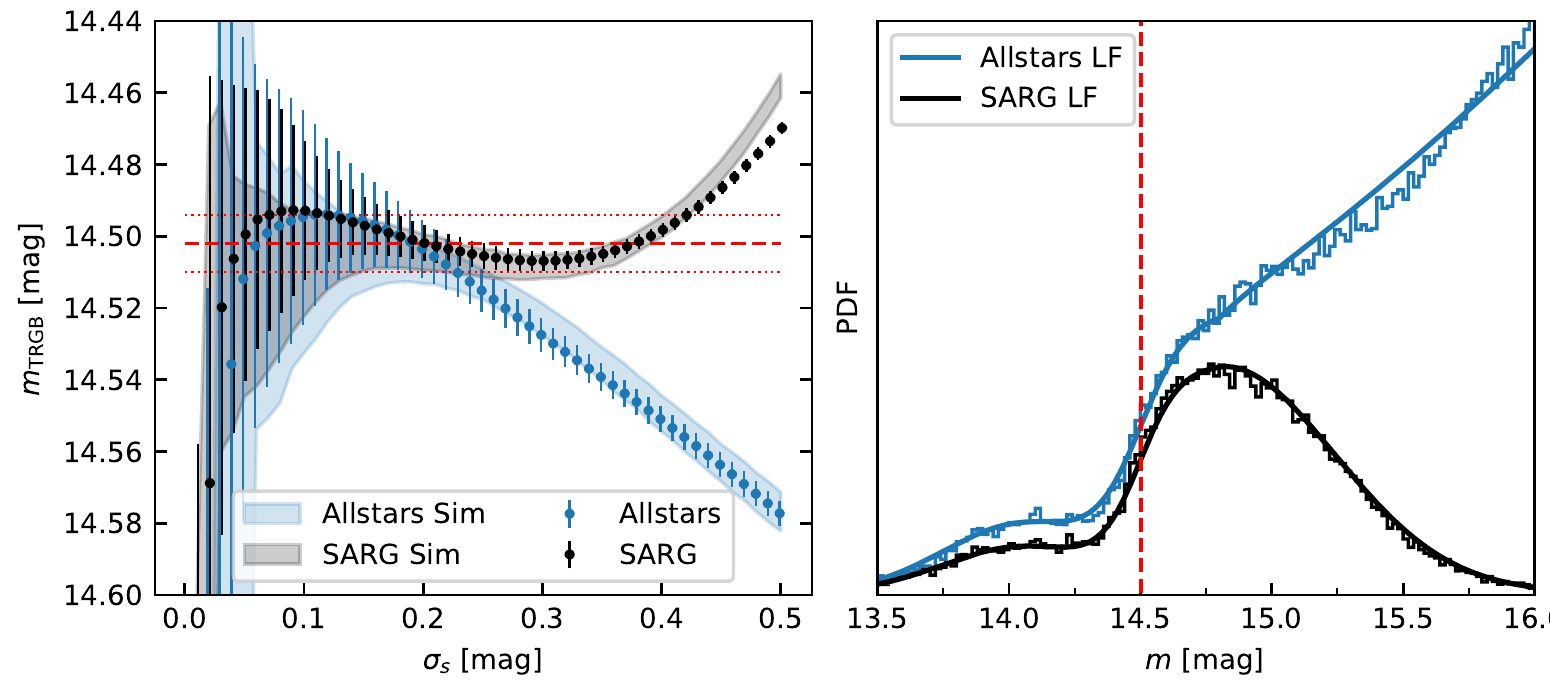}
    \caption{Simulated LFs probe the dependence of  $m_{\mathrm{SARG,mode}}$ on LF shape. {\it Left:} observed (errorbars) and simulated (shaded regions) values of \mt\ against smoothing parameter. The simulations reproduce observed trends at a level of approximately $0.005$\,mag. The simulated true $m_{\mathrm{TRGB,0}} = 14.495$ coincides with the $1\sigma$ interval of our measured TRGB shown as red horizontal lines. {\it Right:} Observed (steps) and simulated (smooth curves) LFs.}
    \label{fig:simLF}
\end{figure}

Figure\,\ref{fig:simLF} illustrates the analytical and observed LFs for the \allstars\ and \varstars\ samples for comparison and compares our empirical \mt\ measurements with simulations for a range of \sigs\ values. We find excellent agreement to within $0.005$\,mag at all \sigs\ levels. In particular, our simulations reproduce the observed smoothing bias.

\begin{figure}
    \centering
    \includegraphics[width=1\textwidth]{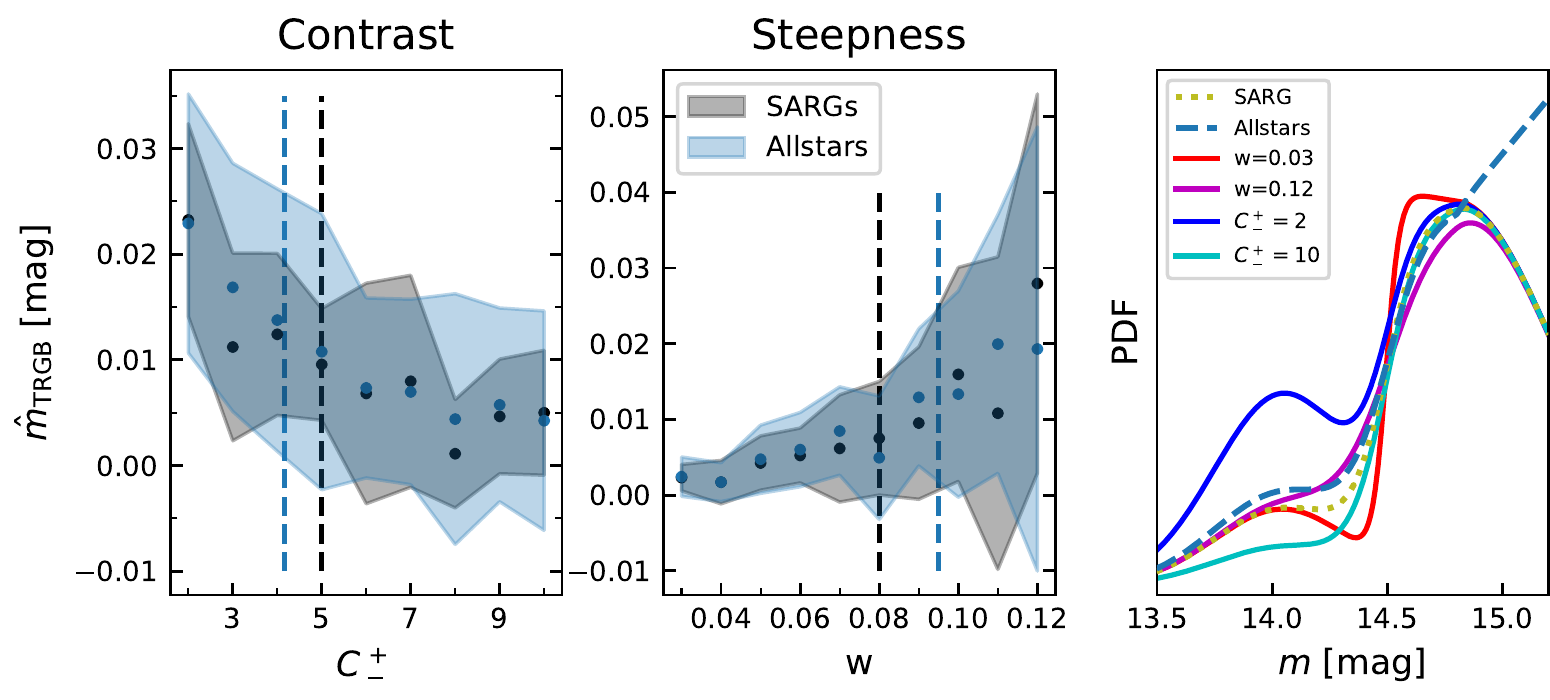}
    \caption{Investigating bias due to TRGB contrast (left) and steepness (center) for \allstars\ (blue) and \varstars\ (black) samples at \sigs$=0.15$\,mag. Equation\,\ref{eq:contrast} defines \Tcont\ independently of \mt. Higher contrast (larger \Tcont) and steeper (smaller $w$) TRGB features reduce bias and enhance precision. Vertical dashed lines indicate the values of the observed samples at \Tcont\ (Eq.\,\ref{eq:contrast}) from Tab.\,\ref{tab:LF_coeffs}. The right-hand panel illustrates how varying these parameter changes the \varstars\ LF.}
    \label{fig:simCont}
\end{figure}

Figure\,\ref{fig:simCont} illustrates the bias $\hat{m} = m_{\mathrm{TRGB}} - m_{\mathrm{TRGB},0}$ due to contrast (\Tcont) and steepness ($w$) for \sigs$=0.15$\,mag and illustrates the corresponding effects on the composite LF. We find smaller bias $\hat{m}$ for higher \Tcont\ values, qualitatively similar to Fig.\,9 in \cite{Wu2022}. We further find that steeper TRGB features tend to be less biased, that the \varstars\ sample exhibits lower dispersion in the simulations than the \allstars\ sample, and that the \varstars\ LF yields slightly lower bias ($\sim 0.009$\,mag) than the \allstars\ LF ($\sim 0.013$\,mag).

\subsection{Sobel Filter Weighting and Tip Contrast Relations\label{sec:TCR}}

Weights ($w$) were introduced to the TRGB method to invoke a notion of statistical significance for the Sobel filter response, $\mathfrak{S}$ \citep{Madore2009}. The edge detection response (EDR) is thus the product  $\mathrm{EDR} = \mathfrak{S} \times w(i)$, and \mt\ is measured as the mode of the EDR. We note that GLOESS smoothing introduces correlation among the very narrow LF bins, complicating the interpretation of weighted EDRs in terms of signal-to-noise. 

Different descriptions of ostensibly identical weighting schemes have been presented in articles associated with the CCHP \Ho\ analysis \citep[e.g.][]{Madore2009,Hatt17,Freedman2019,Hoyt2023}.  
We therefore investigated three weighting options \citep[cf. also][]{Wu2022,Li2023}: a) no weighting, i.e., $w(i)=1$; b) weighting $\mathfrak{S}$ by the inverse Poisson error, i.e., $w(i) = 1/ \sqrt{N[i-1] + N[i+1]}$, as implemented by \cite[their Eq. B1]{Li2023} and described by \cite{Madore2009,Freedman2019,Freedman2021}; c) ``signal-to-noise'' weights $w(i) = ( N[i-1] - N[i+1] ) / \sqrt{N[i-1] + N[i+1]} $ as described in H23, equivalent to weighting $\mathfrak{S}^2$ by the inverse Poisson error.

\begin{figure}
    \centering
\includegraphics{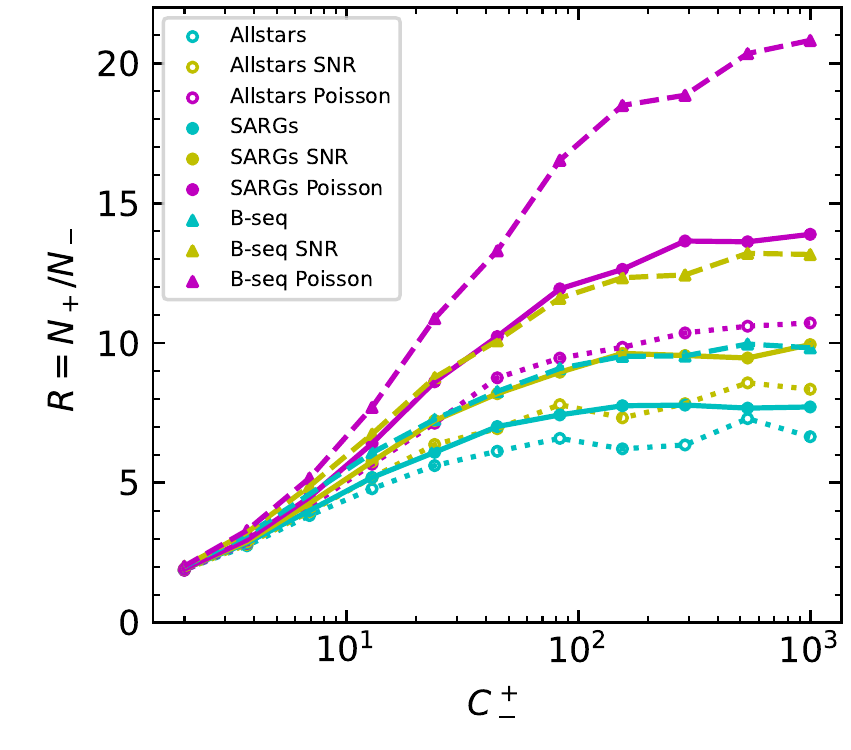}
    \caption{Comparison of input \Tcont\ used to simulate the LFs (cf. Eq.\,\ref{eq:contrast}) and the tip contrast $R$ determined using different samples and EDR weighting schemes. $R$ can differ greatly according to EDR weighting scheme because it is evaluated at increasingly biased \mt. Based on same LFs shown in Fig.\,\ref{fig:EDRTCR}.}
    \label{fig:ConstrastRvsC}
\end{figure}

\begin{figure}
\includegraphics[]{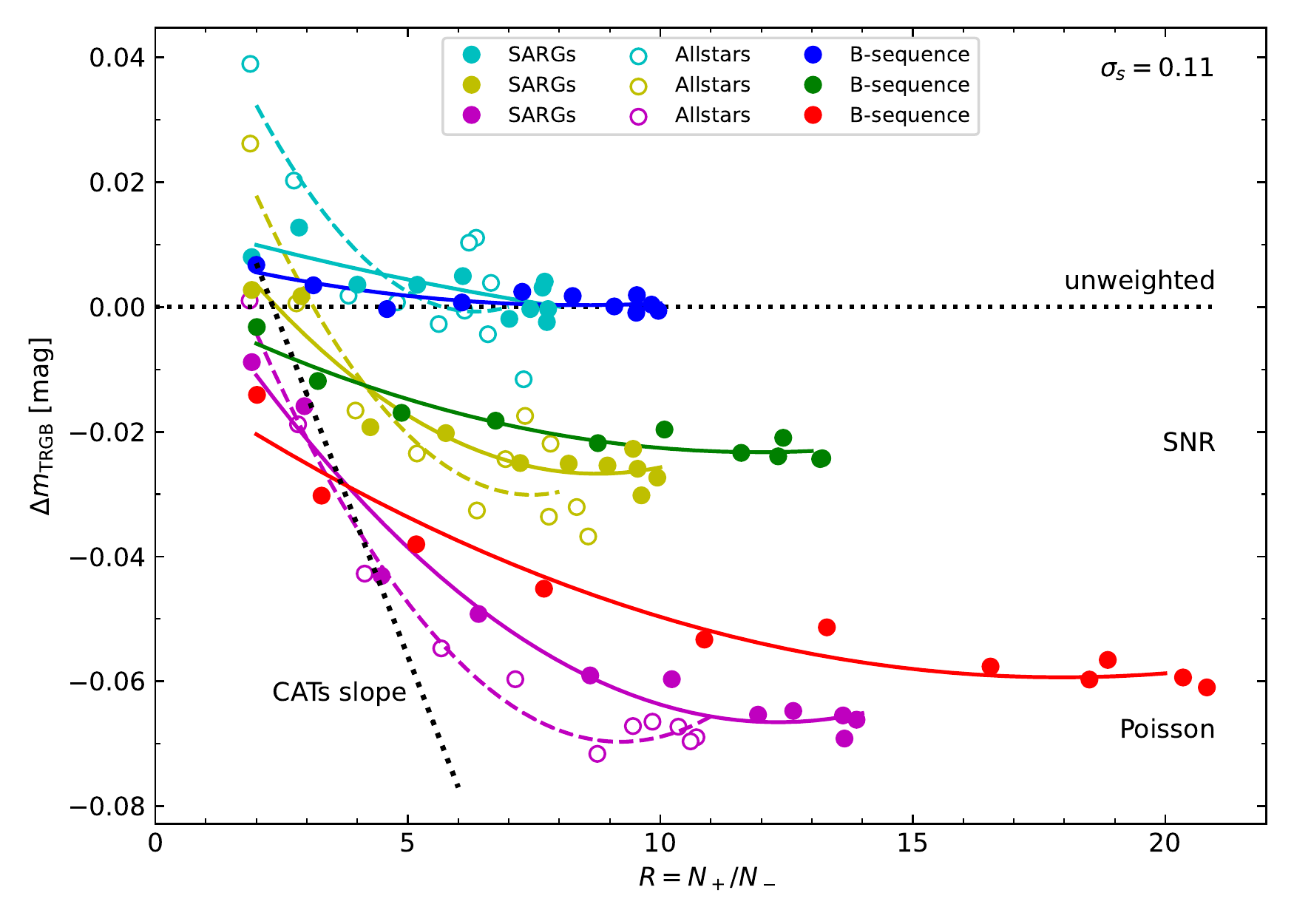}
\caption{Bias induced by weighting Sobel filter responses using different methods. Simulated LFs (\sigs$=0.11$\,mag) for \allstars\ (open circles), \varstars\ (filled cyan, yellow, magenta circles), and \Bseq\ (filled blue, green, and red circles) exhibit consistent bias as a function of tip contrast $R$ when using the same EDR. Regardless of sample, the unweighted EDR yields the least biased, and the Poisson EDR the most biased, results. Differences in $R$ arise from the biased estimation of \mt. All TCRs are non-linear; the linear trend by \cite[black dotted line]{Scolnic2023} is a good approximation for $R$ between $2-5$. Solid lines represent fits to solid symbols, dashed lines those to open symbols. The quadratic fit to the Poisson EDR \allstars\ sample (dashed magenta line) yields $-0.060$\,mag at $R=6.4$.\label{fig:EDRTCR}}
\end{figure}

We simulated \allstars, \varstars, and \Bseq\ sample LFs with different \Tcont\ and computed tip contrast $R = N_+ / N_-$ after determining \mt\ using the different weighting schemes. As shown in Figure\,\ref{fig:ConstrastRvsC}, the resulting $R$ values depend on the sample, precluding a straightforward translation between \Tcont\ and $R$. Figure\,\ref{fig:EDRTCR} illustrates these results and shows that unweighted EDRs (option a) lead to nearly unbiased results above $R \gtrsim 2.5$, and we therefore adopt the unweighted procedure to measure \mt\ empirically (Tab.\,\ref{tab:results}). Option b) yields the most biased results and option c) exhibits intermediate bias. Analogous trends based on synthetic stellar populations are see in Fig.\,9 of \cite{Li2023}. 

Weighted EDRs yield systematically brighter \mt\ values than unweighted EDRs (Fig.\,\ref{fig:EDRTCR}). This, in turn, introduces in a dependence of \mt\ on $R$ (a tip-contrast-relation, TCR) because the number of stars above the tip decreases rapidly towards brighter magnitudes and because $R$ by definition depends on \mt. As a result, the same underlying LF with a given value of \Tcont\ yields different $R$ values depending on EDR weighting scheme (Fig.\,\ref{fig:ConstrastRvsC}). 
Our simulations thus explain the origin of the TCR reported based on observations of SN-host galaxies and LMC subfields \cite{Wu2022,Scolnic2023,Li2023}, and we note the agreement between the linear TCR by \cite{Scolnic2023} and our simulations for $R$ in the range of their LMC subfields ($2 < R < 5$). Weighting $\mathfrak{S}$ of the  observed LFs used in this study yields identical trends and consistent differences among the different weighting options. Studies using weighted EDRs must therefore correct for $R$ differences as done in the CATS analysis \citep{Scolnic2023}, while unweighted EDRs do not require such correction. The quadratic fit to the \allstars\ sample analyzed using Poisson EDR weighting yields a bias of $-0.060$\,mag at $R=6.4$, the contrast of the combined H23 calibration sample.

We choose to report only results based on unweighted EDRs to avoid the bias and complications related to TCRs introduced by weighting. We note that $R$ differences cannot explain the magnitude difference between \mt\ determined using the \ABseq\ samples, in particular for the unweighted EDR used here.

\section{Comparison with Hoyt (2023) \label{sec:hoyt}}

Despite our use of the same \ogle-III footprint, photometry, and \ogle-IV RC reddening maps, our results for \varstars\ and \allstars\ samples differ significantly from \mt\ reported by H23, which was adopted by the CCHP \Ho\ analysis \citep{Freedman2021}. To investigate this difference, we restricted the \allstars\ sample to a color-magnitude range following H23 between two lines: $a = -3\cdot( (V-I)_0 - 1.6 ) + 14.45$ and $b = -3.5 \cdot( (V-I)_0 - 1.95 ) + 14.45$. This color cut dims \mt\ by $0.032$\,mag, cf. Tab.\,\ref{tab:results}, and yields \mio$=14.527$\,mag (labeled \allstars$^\dagger$ in Tab.\,\ref{tab:results}), which is 0.088\,mag \emph{fainter} than $m_I = 14.439$\,mag (Fig.~1 in H23, full \ogle\ footprint). 

We noticed that H23 used an incorrectly low reddening value of $R_I = 1.219$. Correcting $R_I$ to a more reasonable $R_I=1.46$  amplifies the disagreement with H23 by $\sim 0.12 \cdot (1.460 - 1.219)= 0.029$\,mag to $0.117$\,mag. We further note that the statistical uncertainty of $E(V-I)$ values, which are correlated due to the limited resolution of the RC reddening maps, does not reduce as $\sqrt{N}$ as assumed in H23. Additionally, the value of $N$ in this context is arbitrary (H23 used an interval of $0.02$\,mag around \mt). 

\begin{figure}
    \centering
    \includegraphics[width=0.32\textwidth]{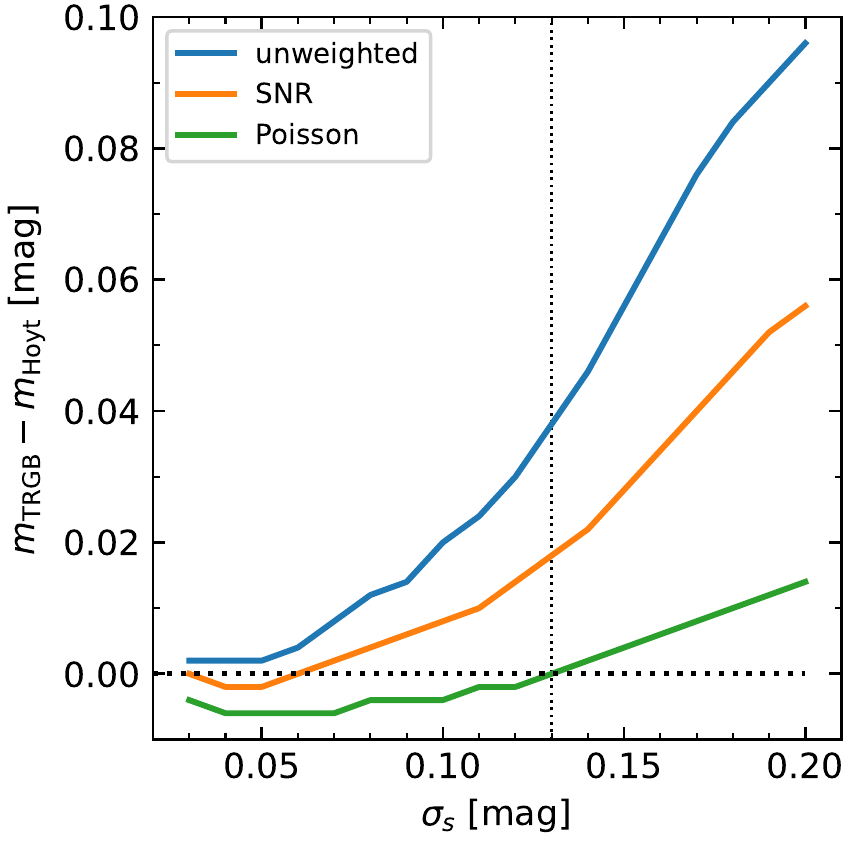}
    \includegraphics[width=0.32\textwidth]{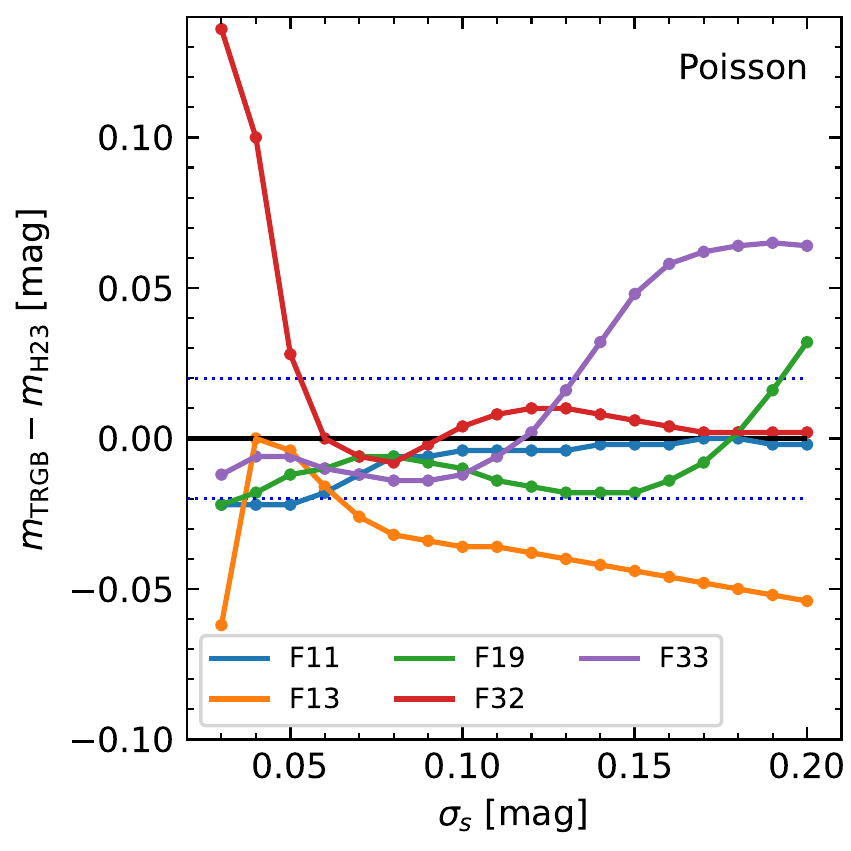}
    \includegraphics[width=0.32\textwidth]{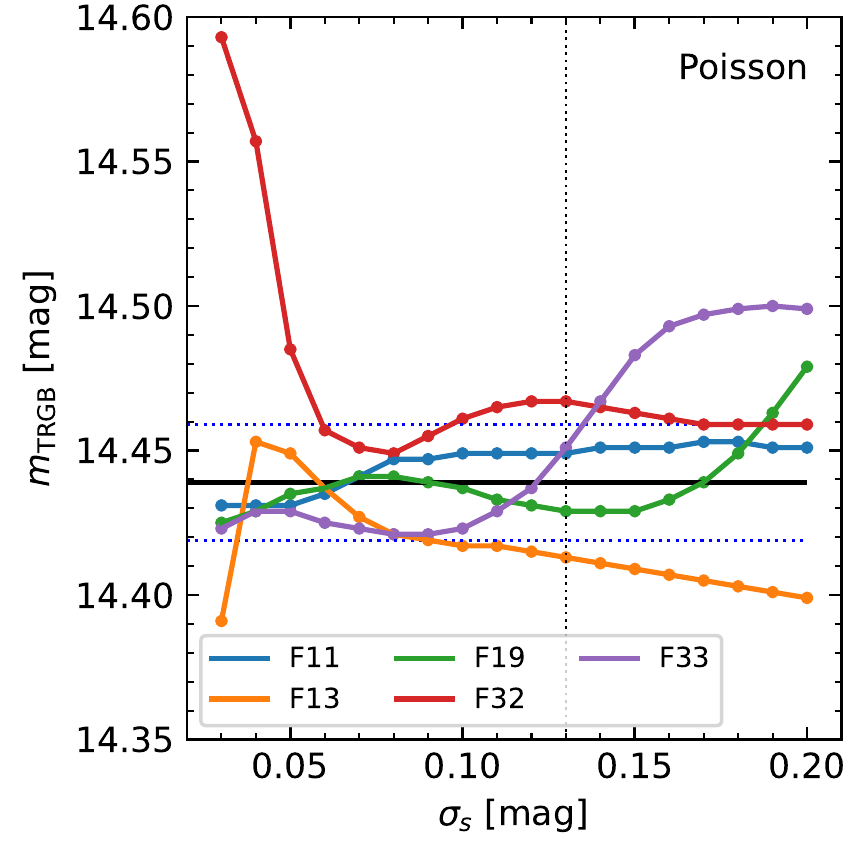}
    \caption{Inferring the EDR weighting scheme and \sigs\ values used by H23. We sought to reproduce both sample selection and methodology explained by H23 (color cut applied, low $R_I=1.219$ used) and measured \mt\ over a range of \sigs\ values using different EDR weighting schemes for the five ``calibration'' fields used by H23 as well as their combination. {\it Left:} Difference of TRGB magnitude minus H23 for three weighting schemes using combined fields. Poisson weighting and \sigs$\approx 0.13$\,mag exactly reproduces H23. {\it Center:} Difference of TRGB magnitude minus H23 for individually selected fields using Poisson weighting, which yields best consistency. No single value of \sigs\ reproduces individual results reported by H23. {\it Right:} \mt\ for the five selected fields using Poisson EDR weighting. Result from combined sample shown as solid black horizontal line for reference together with most likely \sigs\ value used by H23. NB: averaging \mt\ from individual fields does not and is not expected to yield the same \mt\ as the combined set of stars in all fields.}
    \label{fig:HoytSmoothingEDR}
\end{figure}

We traced the origin of this large, $0.117\,$mag, difference to the combination of several systematic differences by simulation and by reproducing H23's analysis step by step, including color restrictions, use of low $R_I$, geometric corrections, and spatial selections. We analyzed the five sub-regions of the \ogle-III footprint H23 labeled ``calibration sample'' (henceforth: F11, F13, F19, F32, and F33), as well as the combined sample of stars located therein (henceforth: H23 sample). Since H23 did not specify which value of \sigs\ was adopted, and given  ambiguous descriptions of the EDR weighting procedure in the body of CCHP articles \citep[e.g.][]{Madore2009,Hatt17,Freedman2019,Hoyt2023}, we computed results for a range of smoothing values $0.02 \le \sigma_s \le 0.2$ and using all three EDR weighting schemes described in App.\,\ref{sec:TCR}. 

Figure\,\ref{fig:HoytSmoothingEDR} shows the comparison of our results with H23's tabulated information. Based on this comparison (left panel in Fig.\,\ref{fig:HoytSmoothingEDR}), we conclude that H23 mostly likely\footnote{Unfortunately, the author was not available for comment despite multiple inquires.} used \sigs$=0.13$\,mag and the Poisson EDR weighting scheme, despite Eq.1 in H23 suggesting that SNR weighting was used. While a combination of lower \sigs\ and SNR weighting can also reproduce H23's results, Poisson weighting yields better field-to-field consistency and is required to reproduce \mt\ close to those reported in their Extended Data Table 2. However, the center panel of Fig.\,\ref{fig:HoytSmoothingEDR} shows that no single value of \sigs\ reproduces the tip magnitude reported in H23, suggesting that \sigs\ was left to vary from field to field. We also note that the CCHP \Ho\ analysis did not specify \sigs\ values used for individual galaxies, apart from the general statement that \sigs\ is typically close to the photometric error. At \sigs$=0.13$\,mag, we find a total range of $\sim 0.05$\,mag variations around the H23 sample result of $14.439$\,mag.

Having thus shown the ability to reproduce H23's result using Poisson EDR weighting and \sigs$\approx 0.13$\,mag, we quantify the origin of the $0.115$\,mag difference with H23 as follows. Incorrect $R_I$ renders \mt\ brighter by $-0.029$\,mag. Smoothing does not significantly contribute to the difference between the H23 calibration sample and our \allstars\ analysis because they used similar \sigs\ values. According to our simulations, EDR weighting brightens \mt\ by approx. $-0.06$\,mag for $R=6.4$ measured in the H23 sample (Fig.\,\ref{fig:EDRTCR}). Geometric corrections brighten \mt\ in the H23 sample by $\sim -0.015$\,mag. The remaining difference of $0.117 -0.029 - 0.06 - 0.015 \approx 0.013$\,mag is thus most likely attributable to the rejection of $\sim 80\%$ of the stars available when considering the H23 calibration sample. According to H23, field selection impacted their \mt\ by merely $0.002$\,mag. However, this could be slightly skewed if a different value of \sigs\ was used. Minor differences in methodology and sample selection, e.g., concerning foreground star cleaning, etc., can easily explain the remaining small difference of $\sim 0.5\%$ in distance.

\begin{figure}
\centering
\includegraphics[width=1\textwidth]{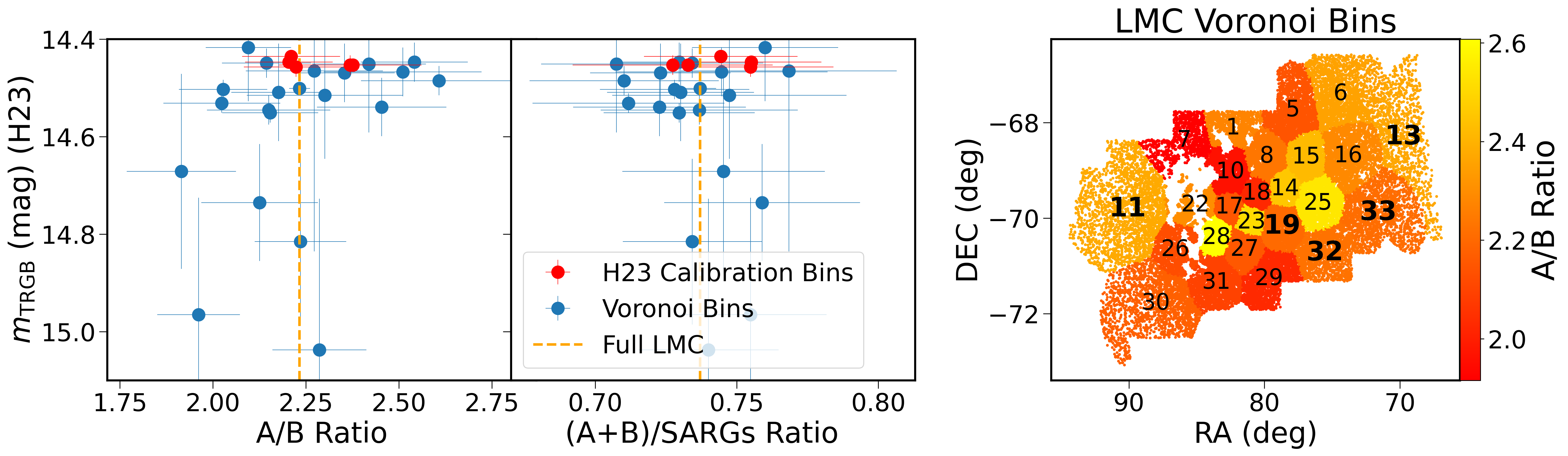}
\caption{The ratio of SARGs on sequences A and B is rather even across the \ogle-III LMC footprint. \textit{Left}: TRGB magnitudes from H23 measured from different spatial following H23 using their Voronoi tessellation versus the ratio of SARGs on sequence A to B. The calibration fields used by H23 are highlighted in red, and the vertical dashed line indicates the mean ratio of A to B. \textit{Center:} same as left but versus the ratio of SARGs on sequences A and B to all \varstars. \textit{Right:} ratio of stars on the \ABseq\ across the LMC \ogle-III footprint with Voronoi spatial bins numbered as in H23 \label{fig:osargspatial}.}
\end{figure}

We note that the H23 calibration sample does not preferentially select the LMC's older RG population. Figure\,\ref{fig:osargspatial} shows that the \Bseq\ stars are located all across the \ogle-III LMC footprint, with only minor variations in the ratio of numbers on sequences A and B. Neither the ratio of the number of RG stars on the A and B sequences, nor the ratio of \ABseq\ stars to all \varstars\ is particularly low in any of the H23 calibration fields. Figure\,\ref{fig:osargspatial} thus shows that spatial selections do not allow one to select old stars in the \ogle-III LMC footprint. However, variability-based selection does allow one to separate old and young RG stars. Importantly, \Bseq\ stars are slightly redder than \Aseq\ stars, in particular close to the RGB tip. The color cut by H23 would tend to preferentially remove these stars, thus removing the brightest old RG stars at the Tip. Color cuts should therefore be considered with great care, in particular in the presence of diverse RG populations.

\bibliography{VariableTRGB}{}
\bibliographystyle{aasjournal}

\end{document}